\documentstyle[prl,aps,multicol]{revtex}
\renewcommand{\narrowtext}{\begin{multicols}{2} \global\columnwidth20.5pc}
\renewcommand{\widetext}{\end{multicols} \global\columnwidth42.5pc}
\input{epsf.tex}

\begin{document}
\bibliographystyle{prsty}%
\draft%
\title{Giant Microwave Absorption in Metallic Grains: Relaxation
  Mechanism}

\author{F. Zhou$^1$, B. Spivak$^1$, N. Taniguchi$^2$~\cite{taniguchi-j}
  and B. L.  Altshuler$^{2,3}$}

\address{$^1$Physics Department, University of Washington, Seattle, WA
  98195\\ $^2$Department of Physics, Massachusetts Institute of Technology,
  Cambridge, MA 02139\\$^3$NEC Research Institute, 4 Independence Way,
  Princeton, NJ 08540}

%\date{\today}

\maketitle

\begin{abstract}
  We show that the low frequency microwave absorption of an ensemble of
  small metallic grains at low temperatures is dominated by a mesoscopic
  relaxation mechanism. Giant positive magnetoresistance and very strong
  temperature dependence of the microwave conductivity is predicted.
\end{abstract}


\narrowtext

Microwave absorption in an ensemble of metallic grains has been
investigated both theoretically and experimentally in many papers. (See
for Refs.~\cite{Gorkov65,Shklovskii82,Perenboom81,Carr84}.)  At high
temperatures quantum effects can be neglected and the effective
microwave conductivity at frequency $\omega$ is given by the classical
Debye formula,
\begin{equation}
  \sigma_{D}=\frac{\omega^2}{\sigma_{cl}}.
\end{equation}
Here $\sigma_{cl}$ is the classical Drude conductivity determined by
elastic scattering.  At low temperatures the quantum nature of electronic
states in grains becomes essential and statistics of the electron levels
determines the microwave
absorption~\cite{Gorkov65,Shklovskii82,Perenboom81,Carr84}.  However all
quantum effects considered so far result in the microwave conductivity of
the order or smaller than $\sigma_{D}$. In this Letter we discuss the
mechanism of microwave absorption which can be much stronger than the
classical one given by Eq.~(1).

At sufficiently high radiation frequency $\omega$ the absorption is
dominated by the resonant mechanism, i.e., by the direct transitions
between electron levels~\cite{Gorkov65,Shklovskii82}, and the microwave
conductivity is determined by the probability density $P(s)$ of energy
spacing $s$ between adjacent levels.  This quantity is usually described
by the Wigner-Dyson distribution. At $s\ll \Delta$, where $\Delta$ is the
mean level spacing, the probability density behaves as~\cite{Mehta} (see
the insert on Fig.~1)
\begin{equation}
  P(s)=C_{\beta} {s^{\beta}}/{\Delta^{\beta+1}}.
\end{equation}
The exponent $\beta$ is determined by the global symmetry of the system.
For the orthogonal ensemble $\beta=1$ and $C_\beta = \pi^2/6$.
In the unitary case, $\beta=2$ and $C_\beta=\pi^2/3$.  Finally, the
symplectic  ensemble is characterized by $\beta=4$ and
$C_\beta=16\pi^2/135$.

It is well known, however, that absorption of radiation in two level
systems is determined at low $\omega$ by the relaxation mechanism.
Mandelstam and Leontovich proposed this mechanism for sound attenuation in
gases with slow internal degrees of freedom~\cite{FluidMechanics}. For
microwave absorption in doped semiconductors relaxation absorption was
suggested by Pollack and Geballe~\cite{Pollak61}.

The contribution of this mechanism to the microwave conductivity of
metallic grains $\sigma_{R}$ depends strongly on the relation between
$\Delta$ and the level broadening $\Gamma$. For $\Gamma\gg \Delta$, the
relaxation mechanism of absorption in granular metals was shown in
Ref.~\cite{Spivak95} to determine under certain conditions both the
microwave conductivity and the magnetoconductance of metallic grains.
Here we consider the opposite limit $\Gamma\ll \Delta$ when the energy
levels are well resolved. It turns out that the microwave
absorption $\sigma_{R}$ as well as the magnetoconductance
are giant in this limit, i.e., $\sigma_R$ exceeds both Eq.~(1) and results
of Ref.~\cite{Spivak95} by several orders of magnitude.

We now discuss the physical picture of the relaxation absorption of
applied electric field of frequency $\omega$.  In adiabatic approximation
temporal energy levels $\epsilon_i(t)$ oscillates with the
same frequency.  Since the populations of the energy levels follows
adiabatically the motion of the levels themselves, the electron
distribution becomes nonequilibrium without any interlevel electron
transitions.  Relaxation of the nonequilibrium distribution due to
inelastic processes leads to the entropy production and therefore to the
absorption of the energy of the external field.

To simplify the situation, we will restrict ourselves to the low
temperature case when $T\ll \Delta$.  (Later we will also discuss the
situation at $\Gamma\ll \Delta<T$ qualitatively.) We will see that at
$\omega\ll T$ the relaxation mechanism of absorption is determined by rare
grains where the first excited level is separated from the ground state by
small energy $s\ll T$. This implies that the ground and the first excited
states in the grains form two level systems which are effectively isolated
from the rest of the spectrum.

The population $N(t)$ of the first excited state with  the energy $s(t)$
is governed by the equation
\begin{equation}
  \frac{dN(t)}{dt}=-\frac{N(t)-N_{0}(s(t))}{\tau_{\epsilon}}.
\label{dN/dt}
\end{equation}
Here $N_{0}(s(t))=(1+\exp ({s(t)}/{T}))^{-1}$ is the adiabatic equilibrium
population of the first excited level and $\tau_{\epsilon}$ is the
relaxation time of the level due to the electron-phonon interaction.
Eq.~(\ref{dN/dt}) is applicable provided $s\gg \Gamma$.

It is crucial for our discussion that the relaxation rate for a two level
system vanishes with $s$.   For a two level system
embedded into 3D insulating environment with phonons, $\tau_{\epsilon}\sim
{s^{-4}}$ for $s\to 0$.  The fact that $\tau_{\epsilon}(s)$ diverges so
rapidly for small $s$ leads, as we will see, to divergences in $\sigma_R$.
To provide the proper cutoff we need to include the
contribution to $\tau_{\epsilon}^{-1}$ from the processes that involve
the second excited state. The latter is typically separated in energy by
$\Delta\gg T$ and we
get a finite, though exponentially small, relaxation rate at $s=0$:
\begin{equation}
\frac{1}{\tau_{\epsilon}(s)}=\frac{1}{\tau_{0}}(\frac{s}{\Delta})^{4}+
\frac{1}{\tau_m},
\label{tau_e}
\end{equation}
where
\begin{equation}
\frac{1}{\tau_0}\sim \frac{\Delta^{5}L^{2}lT }{\Omega_{D}^{2}v^{3}_{s}};\quad
\frac{1}{\tau_m}=\frac{\Delta}{T\tau_0} \exp(-\frac{\Delta}{T}),
\label{tau_0-tau_m}
\end{equation}
$\Omega_D$ is the Debye frequency in the metal, $v_s$ denotes the velocity
of sound, $L$ is the size of the sample and $l$ is the elastic mean free
path. To interpret Eq.~(\ref{tau_0-tau_m}) we present the first term in
Eq.~(\ref{tau_0-tau_m}) in the form
$({s^{3}}/{\Omega_{D}^{2}})({\Delta}/{s})(kl)(kL)^2({T}/{s})$, where
$k=({s}/{\hbar v_{s}})$ is the wave number of a phonon with the energy
$s$. The first factor here corresponds to the conventional expression for
the relaxation rate in clean bulk metal at $\epsilon_i \approx T$.  We are
considering a transition between two particular levels without summing
over the final states (as is usual in the bulk case); this gives rise to
the second factor. In the dirty case ($kl<1$) the electron phonon
relaxation time is known to be suppressed~\cite{Reizer} (third factor).  If
the phonon wave length exceeds the system size $L$ as well, the
electron-phonon coupling in the dipole approximation is reduced, and the
fourth factor appears. Finally at $s<T$ the large population number of
phonons leads to the last factor.

To evaluate $\sigma_{R}$ we consider the power $Q$ absorbed in a grain.
Given the amplitude of the microwave electric field $E$ and the volume of
the grain $V$,
\begin{equation}
  Q=\omega \int dt\frac{ds}{dt} N(t);\qquad \sigma_{R} =\frac{Q}{E^{2}V}.
\end{equation}
In the ohmic approximation we obtain the Debye-type expression for the
averaged microwave conductivity,
\begin{equation}
\sigma_{R}=\langle(\frac{ds}{dE})^2\rangle\frac{1}{V} \int ds P(s)
\frac{dN_0}{ds}\frac{\omega^2\tau_{\epsilon}(s)}
{1+(\omega\tau_{\epsilon}(s))^2}.
\label{def:sigma_R}
\end{equation}
Here $\langle...\rangle$ stands for the averaging over the random
scattering potential. The random matrix elements were assumed to be
statistically independent from the spectrum. This assumption can be
verified by a straightforward calculation. We will present this
calculation elsewhere.

The distribution $P(s)$ of the nearest neighbor spacing $s$ is known to be
determined by the global (spin and/or time reversal) symmetry of the
system.  $\sigma_R$, because of its dependence on $P(s)$ as seen in
Eq.~(\ref{def:sigma_R}), should be very sensitive to weak magnetic field
$H$ and to spin-orbit scattering rate $1/\tau_{so}$. Here we restrict
ourselves only to the three asymptotic regimes: (i) orthogonal, when $H=0$
and $\tau_{so}=\infty$, (ii) unitary, when the magnetic field is strong
and (iii) symplectic ($H=0$, while $\tau_{so}$ is short.) We also discuss
the particularly interesting case of the combination of strong spin-orbit
scattering and  a weak magnetic field. With this exception we do not
present here explicit formulae for the absorption in the crossover
regimes.  Here we will consider only linear absorption.

Substitution of Eq.~(\ref{tau_0-tau_m}) into Eq.~(\ref{def:sigma_R}) gives
\begin{equation}
  \sigma_{R}=\frac{\omega^2 \tau_0}{4V T}\langle (\frac{ds}{dE})^2 \rangle
  \phi(\frac{\Delta}{2T}, \omega\tau_m, \frac{\tau_0}{\tau_m}),
\end{equation}
where
\begin{equation}
\phi(a,b,c)=\int^\infty_0
\frac{(x^4+c)x^{\beta}dx}{(x^4+c)^2+(bc)^2} \frac{1}{\cosh^2(ax)}.
\end{equation}

For orthogonal ($\beta$=1)and unitary ($\beta$=2) ensembles of metallic
grains ($\tau_{so}s^{*}\ll 1$) the typical value of $s$ determining the
integral contribution in Eq.~(7) is
\begin{equation}
  s\sim s^{*} = \Delta (\max[\omega\tau_0, \tau_0/\tau_m])^{1/4},
\end{equation}
provided $s^{*}\ll T,\Delta$.  In this limit, $\sigma_R$ is
\begin{equation}
  \sigma_{R}=\frac{\pi\omega^2\tau_0}{16VT}\langle ( {ds\over
    dE})^2 \rangle (\frac{\tau_{m}}{\tau_0})^{1/2\beta}
  f_{\beta}(\arctan(\omega\tau_m))
\end{equation}
with
\begin{equation}
 f_{\beta}(k)={\rm cosec}(\frac{\pi}{2\beta})(\cos
 k)^{1/2\beta}\cos(\frac{k}{2\beta}).
\end{equation}

The amplitude of electric field decays over Thomas-Fermi screening length
$r_{0}=\sqrt{D/4\pi\sigma_{cl}} \ll L$ from its vacuum value $E$ down to
$E{\omega}/{\sigma_{cl}}\ll E$. (Here $D$ is the diffusion constant of
electrons.) This small electric field in the bulk gives the main
contribution to the classical Debye formula Eq.~(1). On the contrary, the
sensitivity Eq.~(12) is determined by the small region of the width of order
of $r_0$ near the surface, where the value of the electric field is of the
order of $E$.
A calculation analogous to that in Ref.~\cite{Spivak95} gives
\begin{equation}
\langle (\frac{ds}{dE})^2\rangle\sim
\frac{e^{4}r_{0}^{4}}{\beta\sigma_{cl}V}.
\end{equation}

As a result, in the orthogonal case
\begin{equation}
\sigma_{R}^{o}={C\alpha \sigma_{D}} \times
\left\{ \begin{array}{ll}
(\frac{T}{\Delta})^{1/2}\exp (\frac{\Delta}{2T})\ \
\ \ \ & \omega \tau_{m}\ll 1\\ \frac{1}{\sqrt{2}}(\frac{1}
{\omega\tau_0})^{1/2}\ \ \ \ \ & \omega \tau_{m}\gg 1 \end{array} \right. ,
\end{equation}
while for the unitary ensemble,
\begin{equation}
  \sigma_{R}^{u}={C\alpha \sigma_{D}} \times \left \{ \begin{array}{ll}
  (\frac{4T}{\Delta})^{1/4} \exp(\frac{\Delta}{4T}) \ \ \ \ \ & \omega
  \tau_{m} \ll 1 \\ 2 \cos(\frac{\pi}{8})(\frac{1}{\omega\tau_0})^{1/4}\ \
  \ \ \ & \omega \tau_m \gg 1 \end{array} \right. ,
\end{equation}
where
\begin{equation}
\alpha={\pi\Delta^{2}\tau_0}/{372T}
\end{equation}
and $C$ is a constant of order unity which depends on the geometry of the
grain and on the direction of the electric field. If the grain is cubic
and the microwave field is perpendicular to its face,
then $C=1$.
Therefore in both the orthogonal and unitary cases $\sigma_R$ is much
larger than the classical Debye conductivity Eq.~(1).

In the absence of electron-electron Coulomb interaction, a similar approach
is applicable even at $T>\Delta > \Gamma$ when many well resolved levels
participate the absorption.  In this case $\sigma_{R}\sim
\alpha'\sigma_{D}$ and $\alpha'=e^{2}r_{0}^{2}V^{-1}\tau_{\epsilon}(T)$.
This matches the results~\cite{Spivak95} at $\Delta\sim \Gamma$.

As usual, one can drive an orthogonal system into a unitary one by
applying magnetic field $H$.  This leads to a giant negative
magnetoconductance.  Here we consider it only qualitatively and only at
$\omega \tau_M \ll 1$.
The field needed to reduce from $\sigma_R^o$ to $\sigma_R^u$ is
\begin{equation}
H\sim H(s^{*})=(\frac{s^*}{\Delta})^{\frac{1}{2}}H_0,
\end{equation}
where given the cross-section of the grain $A$ and its dimensionless (in
units ${e^2}/{\hbar}$) conductance $g$, the characteristic field $H_0$
is
\begin{equation}
H_{0}=\frac{\hbar c}{eAg^{\frac{1}{2}}}
\end{equation}
The field $H(s^{*})$
reduces $\sigma_{R}$ dramatically from $\sigma^{o}_{R}$ to
$\sigma^{u}_{R}$.

Now let the spin-orbit scattering rate $\tau_{so}^{-1}$ exceed $\Delta$.
In this symplectic case $P(s) \sim s^{4}/\Delta^{5}$ and the integral over
$s$ in Eq.~(6) is determined by $s\sim T$. At low $\omega\ll
\tau_{\epsilon}^{-1}$ (which
means $s^{*}\ll T$) we have
\begin{equation}
  \sigma_{R}^{s}= C{64\pi^2\over 45}\sigma_{D}\alpha \frac{T}{\Delta};
  \quad \mbox{if }\omega \ll ({T}/{\Delta})^4\tau_0^{-1}.
\end{equation}
At small frequencies $\sigma^{s}_R$ turns out to be temperature
independent and larger than $\sigma_D$ by the factor $\alpha{T}/{\Delta}
\sim \tau_0 \Delta \gg 1$.
At high frequencies the condition $s^{*}\ll T$ is violated. In the interval
$T<s^{*}<\Delta$ (or $({T}/{\Delta})^{4}<\omega\tau_{0}<1$) $\sigma_R$
is $\omega$-independent for all ensembles and
\begin{equation}
  \sigma_{R}^{o,u,s}={C C_\beta\over 16\pi \beta (\beta
    +5)}(T/\Delta)^{\beta+3}{t_M \over \tau_0} T
\end{equation}

In a symplectic grain each energy level is double degenerated due to the
T-invariance. An applied magnetic field splits this Kramers doublet and
therefore increases $P(s)$ at small $s$. According to
Kravtsov and Zirnbauer~\cite{Kravtsov92}
\begin{equation}
  P(s,H)-P(s,0)\sim\frac{s^{2}}{s(H)^{3}}\exp\left[-\frac{s^{2}}{s(H)^{2}}
 \right]
\label{Ps-so}
\end{equation}
Therefore the distribution function has a peak at
\begin{equation}
s=s(H)\equiv\Delta\frac{H}{H_0}
\end{equation}
and $P(s)\sim s^2$ below this peak.
Calculation of the sensitivity $\langle (ds/dE)^{2}\rangle$ and
$\tau_{\epsilon}$ for a split Kramers doublet require some caution:
without magnetic field both the external electric field and lattice
deformation are unable to split a doublet (for example $ds/dE=0$ for
$H=0$). Therefore at $H<H_0$
\begin{eqnarray}
  &&\langle (\frac{ds}{dE})^{2}\rangle\sim e^{2}r_0^{2} {\Delta \over
 4\pi\sigma_{cl}}(\frac{H}{H_0})^2,\\
&&\frac{1}{\tau_{\epsilon}}\sim
\frac{1}{\tau_0}(\frac{s}{\Delta})^{4}(\frac{H}{H_0})^{2}+\frac{1}{\tau_m}.
\end{eqnarray}

Substituting Eq.~(\ref{Ps-so}) into Eq.~(\ref{def:sigma_R}) we get the
expression for grain magnetoconductance in the limit
$\omega\tau_{\epsilon}\ll 1$ in the presence of substantial spin-orbit
scattering.  For $s(H)>s^{*}$, i.e., ${H^4}/{H_0^4} > \omega\tau_0,
{\tau_0/\tau_m}$, we get
\begin{equation}
  \sigma_{R}(H)-\sigma_{R}(0)\sim (\frac{H_0}{H})^{3} \sigma^{u}_{R}\gg
  \sigma^{s}(H=0).
\end{equation}
Thus we have a giant positive magnetoconductance that increases rapidly
with the decreasing magnetic field until $H > H_0 {s^* / \Delta}$.
Therefore at $H \sim H_0 {s^* /\Delta}$, the magnetoresistance has a sharp
maximum.
The magnetic field dependence of $\sigma_R$ both for $\tau_{so}\to \infty$
and for short $\tau_{so}$ is qualitatively illustrated in Fig.~1.

Let us now estimate $\sigma_R$.  Consider
a metallic grain with $\Delta \sim 1K$ (the size is about $L \sim 150 {\rm
  \AA}$).  $\tau_{0}$ can be estimated as $\tau_{0} \sim 10^{-7}$ sec.
Then we get $\alpha\sim 10^2$ at $T \sim 0.3$K.
Therefore, at $\omega \sim 10^{5}$Hz we predict $\sigma_R$
in the orthogonal case to be about three orders of magnitude larger than
$\sigma_D$.  Under these conditions $s^* \sim 0.3$K
and $H_0 \sim 10$T (at $g \sim 10$), i.e., $H \sim 3$T will make the
absorption several times smaller.

Note that the magnetoconductance is negative and positive in the
orthogonal and symplectic cases respectively, i.e., the sign of the
magnetoconductance is opposite to usual weak
localization~\cite{Lee85,Altshuler85}. This is natural since classical
microwave conductivity Eq.~(1) is inversely proportional to $\sigma_{cl}$.

So far we have neglected effects of the Coulomb interaction between
electrons. First of all, our consideration was based on the
Wigner-Dyson distribution $P(s)$. Its applicability in the presence of
interactions is not as clear as in their absence.  Experimental study of the
microwave absorption could provide information on this probability
distribution.

Another Coulomb effect to consider is the contribution of the
electron-electron interaction to the level broadening $\Gamma$.  We argue
that in the limit $T\ll \Delta$ this contribution does not exist and
$\Gamma$ is determined entirely by the phonon scattering $\Gamma =
\tau_{\epsilon}^{-1}$.  Coulomb interactions can only renormalize the
one-electron energy of the first excited state but not to broaden it,
because the inelastic transitions due to electron-electron interaction are
forbidden by the energy conservation law. Therefore
Eq.~(\ref{def:sigma_R}) remains valid even for strongly correlated
electrons.

As it has been mentioned above, without Coulomb interaction there is a big
interval $\Gamma \sim \tau_{\epsilon}^{-1} < \Delta < T$ where our two
level approximation is not valid and many levels in each grain contribute
to the absorption. These levels are still well resolved. On the other hand
the electron-electron interaction is known to dominate $\Gamma$ at high
temperatures: $\Gamma_{ee} \geq \Delta$ for $T >
\Delta$~\cite{Altshuler85}.  This makes calculations for $\Delta < T <
\Delta (\tau_{o} \Delta)^{1/5}$ anything but trivial. Indeed, calculated
in Ref.~\cite{Altshuler85} $\Gamma_{ee}$ is the broadening of one-particle
states. The exact many-electron states can be broadened only by phonons.
However, the spacing between these exact states $\Delta_{ex}(\epsilon)$ is
of the order of $\Delta$ only for several low energy excitations. With
increasing the energy $\epsilon$ the spacing $\Delta_{ex}(\epsilon)$
decreases very quickly. We find it possible that the results
of~\cite{Spivak95} are valid until $T>T^{*}$ where $T^{*}$ is determined
by the relation $\Delta_{ex} (T^{*})=\Gamma_{ph}(T^{*})$. If so, the
matching should take place in the interval $\Delta<T<T^{*}$. We have to
admit that much better understanding of the relation between exact states
and quasiparticles and of electron-phonon interaction in the closed system
is needed in order to develop a theory of microwave absorption at
$\Delta<T< \Delta(\tau_{o} \Delta)^{{1}/{5}}$.

In addition to the absorption of the electric field, one can consider the
effect of an a.c. magnetic field.  Here we only mention results of
Ref.~\cite{Kamenev93}.
The authors of these papers assumed that the broadening of the energy
levels is energy independent and found that the energy dissipation in the
time dependent magnetic field can be much larger than the classical one at
$\Delta \gg \Gamma$.  The sign of the magnetoresistance in this case turns
out to be the same as we found here, i.e., magnetoresistance is positive
in an orthogonal ensemble and negative in a symplectic one.
We believe, however, that the energy dependence of $\Gamma$ is crucial for
this problem as well, and we plan to present a quantitative theory
elsewhere.

The authors are grateful to P. Chandra, Y. Galperin, Y. Gefen, M.
Gershenzou, C. Marcus and M. Reizer for useful comments.  The work of F.Z.
and B.S. was supported in part by NSF Grant No. DMR 92-05144, and the work
of N.T. was supported in part by NSF Grant No. DMR 92-14480.

%
% References

%\bibliography{ref}

%
%
% Figure 1
\begin{figure}
%\vspace{1cm}
\centerline{\epsfxsize=8cm \epsfbox{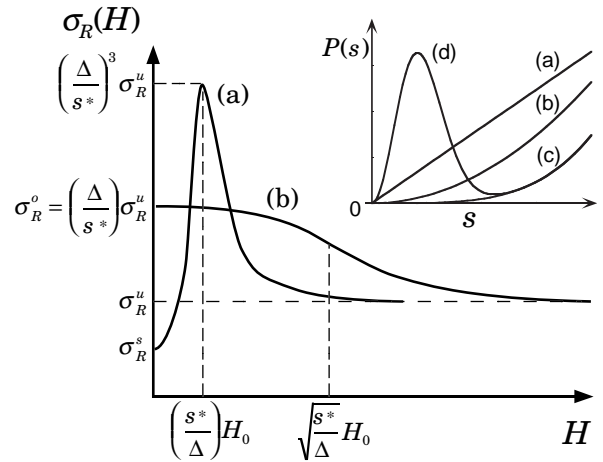}}
\caption{Magnetic field dependence of the microwave conductivity at
  (a) strong spin-orbit scattering, and (b) $\tau_{so}\to \infty$.  Note a
  sharp and high peak in the first case due to the splitting of the
  Kramers doublets.  Parameters $\sigma_R^{o,u,s}$, $s^*$, $\Delta$
  and $H_0$ are determined in the text.  Insert: $P(s)$ at small $s$
  ($s\ll \Delta$): (1) orthogonal, (2) unitary, (3) symplectic (4)
  $\tau_{so}\Delta >1$ and finite $H$. }
\label{fig1}
\end{figure}

\widetext
\end{document}